\documentclass[12pt,a4paper]{article}
\usepackage{epsfig}
\title{Collapse of 4D  random geometries}
\author{P.~Bialas \and Z.~Burda}
\begin{document}
\maketitle
\begin{abstract}
We extend the analysis of the Backgammon model to an 
ensemble with a fixed number of balls and a fluctuating 
number of boxes. In this ensemble the model exhibits 
a first order phase transition analogous to the one in higher 
dimensional simplicial gravity. The transition relies on 
a kinematic condensation and reflects a crisis of the 
integration measure which is probably a part of the 
more general problem with the measure for
functional integration over higher ($d>2$) 
dimensional Riemannian structures. 

\end{abstract}


Recently we have proposed a scenario for the phase transition 
in higher dimensional simplicial quantum gravity 
based on the balls-in-boxes model \cite{bbpt,bbj} also called 
Backgammon model in a different context \cite{rit}.
It has been established that simplicial gravity has two phases:
crumpled  and elongated. In the elongated phase it is effectively described by 
the Branched Polymer model \cite{aj}. In the crumpled phase is characterized by the
appearance of a singular vertices  which gather around them an extensive part
of the volume \cite{jap,ckr}. The transition between those two phases is of 
first order \cite{bbkp,b}.
The constrained-mean-field interpretation based on Backgammon model captures
all those features except the order of the transition. 

In this letter we show that the modified balls-in-boxes model
undergoes a first order phase transition of the same type as in 
simplicial gravity. 


We start with a brief review of the properties the 
Backgammon model. The model describes $N$ balls 
distributed in $M$ boxes. The 
partition function reads~\cite{bbj,rit}: 
\begin{eqnarray}
Z_{M,N}= 
\sum_{q_1,\ldots,q_M}p(q_1)\cdots p(q_M)\delta_{q_1+\cdots+q_M,N}
\label{zmn}
\end{eqnarray}
The numbers of balls $q_i$ and $q_j$ in two boxes $i$ and $j$
are independent as reflected by the 
factorization of the total thermodynamic weight to
one-box weights $p(q)$'s. The factorization is weakly
broken by the constrain on the total number of balls.
This constrain introduces  correlations
between boxes and may result  in
the condensation of balls into one box \cite{bbj}.

We are interested in the limit of $N\rightarrow\infty$ and $r = M/N$
fixed. We call the 
quantity $r$ curvature. In this limit one expects the partition function to behave
as\footnote{One might alternatively formulate the limit in terms of
$M$ and $\rho=N/M$~: $Z_{M,N} \sim \exp\left(M f(\rho)+ \cdots\right)$ as in
\cite{bbj}.  In fact this would look more natural, since $\rho$ has an
interpretation of the average number of balls per box and not the
number of boxes per ball as defined by $r$. Moreover it
would be more natural to keep $M$ in front of $f$ since then $f$ had
an interpretation of the free energy density per box. We use the
convention with $r$ and $N$ on purpose since it is better suited for
calculations in the ensemble with varying $M$ which is the
subject of the letter.}
\begin{eqnarray}
Z_{M,N} \sim \exp\left( N f(r)+ \cdots\right)
\label{rN}
\end{eqnarray}
where the dots stand for corrections growing slower 
than linearly in $N$.
The function $f(r)$ is the free energy density per ball.
The quantity $r$ is bounded from below by $1/N$ since
the system has to have at least one box. Additionally to fix attention 
we introduce an upper bound on $r$
by requiring that each box must contain at least one ball. 
These give together~: $1/N \le r \le 1$.

The phase structure 
of the model is encoded in the analytic properties of the
function (the $\log$ does not change those properties and is 
added for convenience):
\begin{eqnarray}
K(\mu)= -\log \sum_q p(q) e^{-\mu q}
\label{km}
\end{eqnarray}
In particular if the series under the logarithm has a finite radius of 
convergence then the model given by the partition function 
(\ref{zmn}) has a very interesting phase structure \cite{bbj,bb}.
Denote the real value of $\mu$ at the radius of convergence by $\mu_{cr}$, and restrict
ourselves to the situation when the value of the
derivative $K'(\mu_{cr})$ is finite as happens for instance
for the weights of the form $p(q) = q^{-\alpha} e^{-\mu_{cr}q}$
when $\alpha>2$. In this case it can be shown that the system
has two phases depending on the value of curvature $r$
\cite{bbj}. 

The free energy density $f(r)$ (\ref{rN}) can be found by the 
steepest descent method. One obtains the following 
expression~:
\begin{eqnarray}
f(r)  = \left\{ \begin{array}{lll} - r K(\mu_*(r)) + \mu_*(r)
& \mbox{for} & r>r_{cr}  \\ && \\
- r \kappa_{cr} + \mu_{cr} & \mbox{for} & r \le r_{cr} \end{array} \right.
\label{fr}
\end{eqnarray}
where $\mu_*(r)$ is given by the solution of the saddle
point equation
\begin{eqnarray}
\frac{1}{r} = K'(\mu_*)
\label{sp}
\end{eqnarray}
The positivity of the weights $p(q)$'s implies that the
function on the right hand side of (\ref{sp}) is monotonic.
From this follows that the solution $\mu_*$ is 
unique and that it is a monotonic real function of $r$.
For $r=1$ it is infinite. When $r$ decreases so does $\mu_*$.
The equation has a solution for $\mu_*$ as long as $r$ is 
larger then a critical value $r_{cr}$. The critical value
$r_{cr}$ is obtained from the equation (\ref{sp}) when
on the right hand side $\mu_*=\mu_{cr}$ reaches the singularity
of $K$. At the critical value the free energy is given
by the formula in the lower line of (\ref{fr})
where $\kappa_{cr} = K(\mu_{cr})$ and sticks to this  value even
when one $r$ is lower since $\mu_*$ may not move beyond $\mu_{cr}$. At this point the system undergoes a continuous 
phase transition \cite{bbj}.
 
In simplicial gravity the analog of the quantity $r$ is the ratio
$N_0/N_4$ where the $N_0$ is the number of vertices and $N_4$ is the
number of 4-simplices.  The vertices play the role of boxes and
the 4-simplices that of balls. The number of balls in one box is analogue of the
number of vertices sharing one vertex {\em ie} vertex order. The sum
of vertex orders $q_i$ is constrained by the relation: $q_1
+\cdots+q_{N_0}=5N_4$. The ratio $N_0/N_4$ is a linear function of
Regge curvature.  The average curvature is varied through a coupling to
the number of vertices.

In this letter we study the model with fluctuating number of boxes which
more closely resembles the simplicial  gravity scenario. The partition function of
this model is :
\begin{eqnarray}
Z(\kappa,N)=\sum_{M=1}^N Z_{M,N}e^{\kappa M}
\end{eqnarray}
Substituting $Z_{M,N}$ by (\ref{rN}) we get in
the large $N$ limit~:
\begin{eqnarray}\label{Zk}
Z(\kappa,N) = N \int_0^1\!\!\mbox{d}r \exp N \Big( \kappa r + f(r) \Big)
\end{eqnarray}
where $f(r)$ has the form (\ref{fr}). 
The integrand is the distribution
of the curvature for a given $\kappa$. For large $N$
one expects the integrand to approach gaussian shape 
centered around $r_*$ being the solution 
of the saddle point equation~:
\begin{eqnarray}
\kappa + f'(r_*) = 0
\label{fprim}
\end{eqnarray}
For $\kappa > \kappa_{cr}$ this equation
reduces to $\kappa =  K(\mu_{sp}(r_{*}))$ 
which has a unique solution for $r_*$. 
The value of $r_*$ being the position of the gaussian distribution
is the average curvature  
in the limit $N\rightarrow \infty$. 
This situation continues as long as $\kappa > \kappa_{cr}$.

Below the critical point $\kappa < \kappa_{cr}$ the free energy $f(r)$ is
linear in $r$ and the saddle point equation (\ref{fprim}) has no
solution. The integrand is not gaussian anymore but a decreasing
function of $r$. It is exponential~: $\exp N(\kappa - \kappa_{cr})r$
for $r<r_{cr}$.  For large $N$ only this exponential part matters in the
integral (\ref{Zk}) and one finds that $<r>\sim1/N$. 

Exactly at the critical point $\kappa=\kappa_{cr}$,
the equation (\ref{fprim}) is fulfilled by all $r$ between
$0$ and $r_{cr}$ as can be seen from the second line of (\ref{fr}).
The intuition behind this is that 
the curvature $r$  must abruptly move between two separated regimes, 
namely gaussian at $r=r_{cr}$ and exponential at $r=0$.
Therefore $r$ stays undetermined by the saddle point equation.
To fix the shape of the integrand one has to consider finite 
$N$ systems.

\begin{figure}[t]
\begin{center}
\epsfig{file=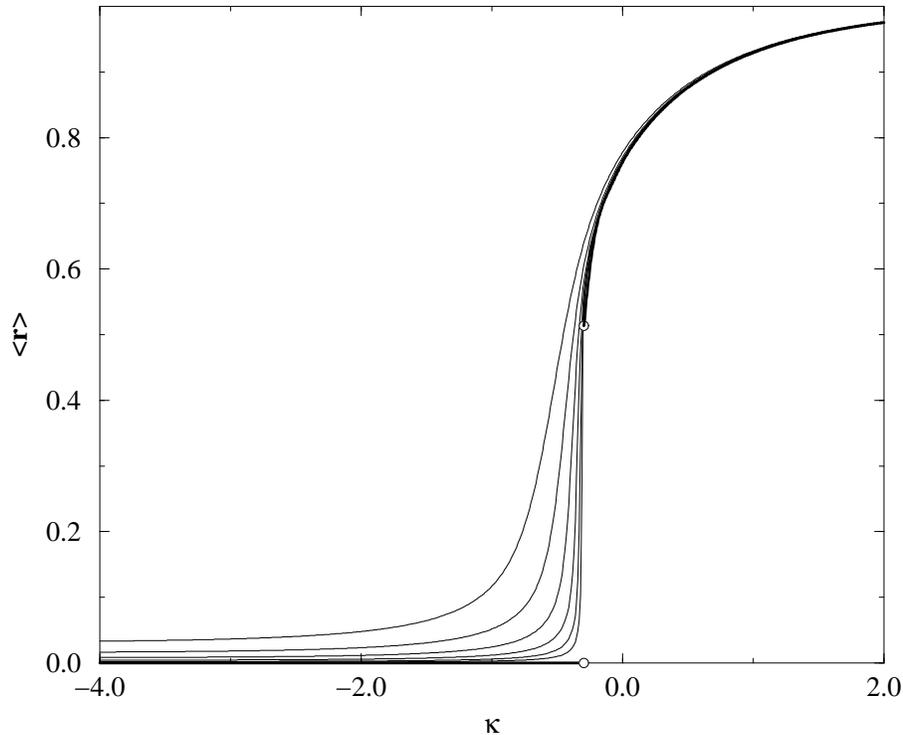,width=14cm,bbllx=12,bblly=50,bburx=576,bbury=450,clip=true}
\end{center}
\caption{Average curvature as the function of $\kappa$ for infinite N (bold line)  32, 64, 128, 256, 512 and 1024 balls}
\end{figure}
In the figure 1 we show average curvature 
$\langle r \rangle$ as a function of $\kappa$ for weights $p(q)=q^{-\frac{5}{2}}$. For 
$\kappa > \kappa_{cr}$ it is a solution of the saddle point
equation (\ref{fprim}). It stops at $r_{cr}$ and falls
to zero. In the same figure we plot average curvature
for some finite values of $N$. The results were obtained
by a recursive technique described in \cite{bbj}. 
In the gaussian phase $\kappa > \kappa_{cr}$ the curves lie 
close to each other indicating that the 
finite size effects are small there.
For large $\kappa$ the curves approach asymptotically the 
upper kinematic bound $r=1$.
In the exponential (kinematic) phase $\kappa < \kappa_{cr}$ 
finite size effects are stronger reflecting the size dependence
of the lower kinematic bound $1/N$. This also agrees with the results
of numerical simulations of simplicial gravity, but there
the kinematic bound is $1/\sqrt{N}$ rather than  $1/N$.

The steepest part of the curves corresponds to the 
pseudocritical region.  
It is a crossover region where the two phases coexist.
The spread of the pseudocritcial region is
$\delta \kappa \sim 1/N$ and reduces to one point for
infinite $N$. In the figure 2 we show the distributions of $r$
for $N=512$ and $N=1024$. One sees coexistence of two phases.
The fraction of either phase depends on a particular choice
of $\kappa$ in the pseudocritical region.  We picked its 
values in such a way as to have more or less the same heights of both the
peaks. The similar two peak structure has been observed in simplicial
gravity for systems with 32000 and 64000 simplices \cite{bbkp,b} 
\begin{figure}[t]
\begin{center}
\epsfig{file=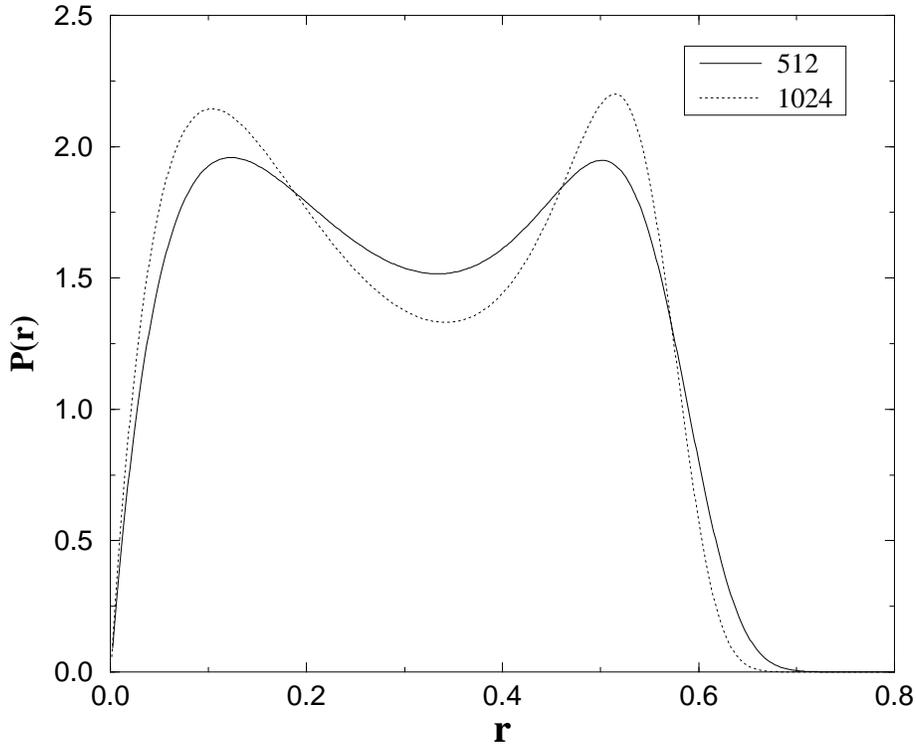,width=14cm,bbllx=12,bblly=50,bburx=576,bbury=450,clip=true}
\end{center}
\caption{Distribution of the curvature for 512 ($\kappa=-0.32184$)
  and 1024 ($\kappa=-0.31910$)  near  phase transition }
\end{figure}

This nicely extends the results of \cite{bbpt} showing that all the
thermodynamical (non-geometrical) features of the simplicial gravity
can be explained by this very simple model.  It seems that the mean
field sets in already for $d\ge3$.  The phase transition from Branched
Polymer phase to the crumpled phase is kinematical {\em ie} it is
associated with system approaching the phase space boundary
rather then diverging correlations (interactions). We think that this 
reflects the crisis of the measure in higher dimensional geometries. 
However it may also be that the gravity does not exist alone and
needs some matter fields which 
would locally smooth the manifold and open a physical
window. Unfortunately, fermions on random lattice
seem to be for the time being far beyond numerical access.

Of course the presented model can only capture the bulk 
features of the system as all the geometry has been  integrated out,
so its simplicity does not exclude the possibility of
interesting geometrical behavior.
One simple  geometrical realization of Bacgammon is BP model \cite{bbj}.
In this model exactly at the phase transition
system enters a third phase which has a different Hausdorff dimension \cite{bb,jk}.
It may be that the same phenomenon happens in simplicial gravity. 

The authors thanks Desmond Johnston and Jerzy Jurkiewicz for helpful
comments and discussion.  This work was partially supported by KBN
grants 2P03B19609.


\begin{thebibliography}{99}
\bibitem{bbpt} P.~Bialas, Z.~Burda, B.~Petersson, J.~Tabaczek
Nucl. Phys. {\bf B495} (1997) 463.  
\bibitem{bbj} P.~Bialas, Z.~Burda, D.~Johnston Nucl. Phys. {\bf B493}
(1997) 505.
\bibitem{rit} F. Ritort, Phys. Rev. Lett. {\bf 75} (1995) 1190\\
               F. Ritort and S. Franz, Europhys Lett. {\bf 31} (1995) 507,
[cond-mat/9505115];\\
               F. Ritort and S. Franz, ``Glassy mean field behaviour of the backgammon model'',
[cond-mat/9508133];\\
                C. Godreche, J.P. Bouchaud and M. Mezard,
 J. Phys {\bf A28}, (1995) L603; \\
               C. Godreche and J.M. Luck, J. Phys {\bf A29} (1996) 1915.
\bibitem{aj}  J.~Ambj\o rn, J.~Jurkiewicz, Nucl. Phys. {\bf B541} (1995) 643. 
\bibitem{jap} T.Hotta, T. Izubuchi, J.~Nishimura, Nucl. Phys {\bf B}
(Proc. Suppl.) {\bf 47} (1995) 609. 
\bibitem{ckr} S.~Cateral, J.~Kogut, R.~Renken, G.~Thorleifsson,
Nucl. Phys. {\bf B468} (1996) 263. 
\bibitem{bbkp} P.~Bialas, Z.~Burda, A.~Krzywicki, B.~Petersson,
Nucl. Phys. {\bf B472}, (1996) 293.  
\bibitem{b}   B.~V.~de~Bakker Phys. Lett. {\bf B389} (1996) 238.
\bibitem{bb}  P.~Bialas, Z.~Burda, Phys. Lett. {\bf B384}(1996) 75.  
\bibitem{jk}  J.~Jurkiewicz, A.~Krzywicki Phys. Lett. {\bf B392}(1997) 291.
\end{thebibliography}
\end{document}